\documentclass[english,a4paper,pre,twocolumn,amsmath,amssymb,longbibliography]{revtex4-1}
\usepackage{bm}
\usepackage{graphicx,color}
\usepackage{soul}
\usepackage[colorlinks,linkcolor=magenta,citecolor=magenta]{hyperref}

\newcommand{\dd}{\mathrm{d}}

\newcommand{\mean}[1]{\langle #1 \rangle}

\newcommand{\IInt}[3]{\int_{#2}^{#3}\dd #1\;}

\newcommand{\argmax}{\mathop{\mathrm{argmax}}\limits}

\renewcommand{\vec}[1]{\mathbf #1}

\newcommand{\eps}{\varepsilon}

\newcommand{\lam}{\lambda}
\newcommand{\vhi}{\varphi}
\newcommand{\sig}{\sigma}
\newcommand{\om}{\omega}

\newcommand{\x}{\vec r}
\newcommand{\tx}{\tau_\text{r}}
\newcommand{\nois}{\bm\xi}

\begin{document}
	
\title{Hunting active Brownian particles: Learning optimal behavior}

\author{Marcel Gerhard}
\author{Ashreya Jayaram}
\author{Andreas Fischer}
\author{Thomas Speck}
\affiliation{Institut f\"ur Physik, Johannes Gutenberg-Universit\"at Mainz, Staudingerweg 7-9, 55128 Mainz, Germany}

\begin{abstract}
	We numerically study active Brownian particles that can respond to environmental cues through a small set of actions (switching their motility and turning left or right with respect to some direction) which are motivated by recent experiments with colloidal self-propelled Janus particles. We employ reinforcement learning to find optimal mappings between the state of particles and these actions. Specifically, we first consider a predator-prey situation in which prey particles try to avoid a predator. Using as reward the squared distance from the predator, we discuss the merits of three state-action sets and show that turning away from the predator is the most successful strategy. We then remove the predator and employ as collective reward the local concentration of signaling molecules exuded by all particles and show that aligning with the concentration gradient leads to chemotactic collapse into a single cluster. Our results illustrate a promising route to obtain local interaction rules and design collective states in active matter.
\end{abstract}

\maketitle


\section{Introduction}

The defining feature of microscopic active particles is their persistent motion: in contrast to passive diffusion they carry an orientation along which displacements are more likely. Such persistent motion is exhibited by bacteria propelled by flagella~\cite{elgeti15} and synthetic colloidal particles that exploit phoretic mechanisms to propel themselves through self-generated gradients~\cite{bech16}. The interplay of this self-propulsion with excluded volume and particle shape leads to a range of collective dynamic states~\cite{roadmap} such as dense clusters, swarms, and flocks~\cite{vicsek12}. Given the wealth of possible collective states emerging already from minimal models, data-driven modeling has attracted interest recently~\cite{cichos20,dulaney20}.

What sets living active matter apart from synthetic active matter is the \emph{intrinsic} ability to respond to external stimuli beyond the direct physical forces generated by gradients. Bacteria and other microorganisms can sense their environment and, \emph{inter alia}, adapt their motility~\cite{wilde17}. For example, some bacteria organize into biofilms~\cite{parsek05} through quorum sensing~\cite{brown01}, \emph{i.e.}, they respond to the concentration of certain signaling molecules exuded by all members of the population. Transferring similar capabilities to synthetic active matter opens the route to program novel collective behavior with the outlook to perform useful tasks~\cite{koumakis13,rubenstein14,sitti15}. Ultimately, this requires internal degrees of freedom computing a response~\cite{horsman14,nava18}. As an intermediate step to elucidate the basic principles, this computation could be performed by an external agent which then acts back on the system.

Recent advances of feedback techniques have demonstrated exquisite control over self-propelled colloidal particles allowing to implement interaction rules that go beyond steric volume exclusion and phoretic and hydrodynamic coupling. Janus particles propelled through the local phase separation of a binary solvent can be addressed individually to implement motility switching~\cite{haufle16,bauerle18,fischer20} and perception-based interactions~\cite{lavergne19}. Active dimers can adapt the propulsion speed depending on their orientation~\cite{sprenger20}. Local heating of a colloidal particle can be exploited for steering~\cite{qian13,bregulla14}, pattern formation~\cite{khadka18}, motility control~\cite{soker21}, and recently has been combined with reinforcement learning to guide a single particle towards a target side~\cite{muinos21}. Reinforcement learning together with computer simulations has been employed to navigate flows~\cite{gazzola16,colabrese17,verma18}, to control shape deformations of microswimmers~\cite{tsang20,hartl21}, to induce flocking of active particles~\cite{durve20,falk21}, and to steer a single active particle through an external potential~\cite{schneider19,liebchen19}.

In this work, we further explore reinforcement learning~\cite{reinforcement} to determine optimal single-particle actions in response to ``states'' representing the information that is available to the particle. {The policy which maximizes the reward is identified as the optimal behavior. We consider a small set of actions that have been implemented experimentally for colloidal Janus particles: turning motility on or off~\cite{bauerle18}, and exerting a torque that makes the particle turn left or right~\cite{lozano16}. The particle dynamics is that of active Brownian particles with propulsion speed and torque depending on the chosen action. As a concrete illustration, we study a simple predator-prey system~\cite{sengupta11,schwarzl16} with an absorbing boundary (the ``predator''), which induces a particle current. Avoiding a predator is complementary to optimal search (and foraging) strategies, which have received intensive theoretical scrutiny~\cite{benichou05,benichou11,bartumeus14,schwarz16,kromer20}. Using as reward the distance to the predator, we evaluate three state-action sets leading to different currents. Finally, we apply reinforcement learning to an interacting suspension of active Brownian particles, demonstrating that it leads to chemotactic collapse for sufficiently large torques overcoming rotational diffusion~\cite{theurkauff12,pohl14}.


\section{Predator-prey system}

We first consider an ideal gas of $N$ non-interacting particles moving in two dimensions in the presence of a single ``predator''. To conserve density, whenever a ``prey'' particle comes within distance $a$ of the predator it is removed and placed at a random position within the system. This corresponds to an absorbing circular boundary at $r=a$ inducing a total current $J$ leaving the system. The accessible area is $A=L^2-\pi a^2$ excluding the disc around the predator with global density $\bar\rho=N/A$. In the simulations, we employ a square box with edge length $L$ employing periodic boundaries, whereas in the analytical calculations we place the predator at the origin. All numerical results are reported employing the length parameter $\sig$ as unit of length and $\sig^2/D_0$ as unit of time with bare diffusion coefficient $D_0$. For non-interacting particles, $\sig$ still determines the effective particle size. Throughout, the radius of the absorbing circular boundary is $a=1.66\sigma$.

\subsection{Passive diffusion}

\begin{figure}[b!]
	\centering
	\includegraphics{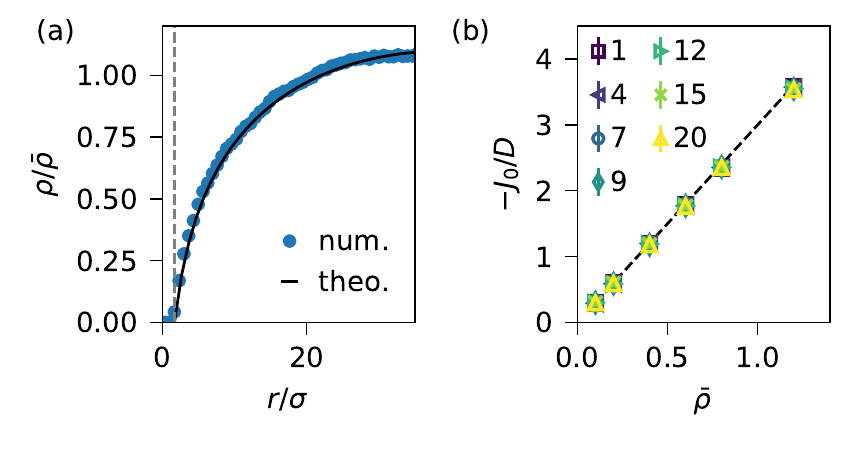}
	\caption{Passive diffusion ($L=70$). (a)~Density profile $\rho(r)$ away from the predator with absorbing boundary at $r=a$ (dashed line) for effective diffusion coefficient $D=1$ in a square system. (b)~Current $-J_0$ (corresponding to the prey caught per time) as a function of global density $\bar\rho$ for several diffusion coefficients $D$ (see legend for values). The dashed line is the prediction Eq.~\eqref{eq:J:p} for $-J_0/D$.}
	\label{fig:passive}
\end{figure}

As reference, we first consider passive diffusion of prey particles with diffusion coefficient $D_0$. The current is $\vec j=-D\nabla\rho$, where $\rho$ is the number density of prey particles. Here we have shifted coordinates so that in the following the predator is at the origin and its diffusive motion enters through the (dimensionless) effective diffusion coefficient $D=1+D_\text{pred}$. In steady state, $\vec j=j(r)\vec e_r$ with $r$ the distance from the predator and $\vec e_r$ the radial unit vector. The total current leaving the system is $J=2\pi aj(a)<0$. The diffusion equation becomes $D\nabla^2\rho+s=0$ with $s=-J/A>0$ the uniform rate at which removed particles respawn. The solution reads
\begin{equation}
  \rho(r) = b\ln(r/a) - \frac{s}{4D}(r^2-a^2)
  \label{eq:rho:p}
\end{equation}
with boundary condition $\rho(a)=0$ and integration constant $b\propto\bar\rho$ determined by the conservation of the total number of prey particles. The radial current is $j(r)=-D\rho'(r)=-Db/r+sr/2$, whereby the prime denotes the derivative with respect to $r$. The total current for passive diffusion thus becomes
\begin{equation}
  J_0 = -2\pi Db\left(1+\frac{\pi a^2}{A}\right)^{-1}.
  \label{eq:J:p}
\end{equation}
Since $s\propto D$ is proportional to the current and thus to the diffusion coefficient, the density profile Eq.~\eqref{eq:rho:p} is independent of $D$. In Fig.~\ref{fig:passive}(a), we show the numerically obtained density profile $\rho(r)$ in a square system with periodic boundaries together with Eq.~\eqref{eq:rho:p}. To this end, the equations of motion are integrated with time step $\delta=1.5\times10^{-5}$ employing the Euler-Maruyama scheme. The negative current $-J_0/D>0$ is plotted in Fig.~\ref{fig:passive}(b) and shows the expected linear increase with the global density.

\subsection{Free active diffusion}

In the next step, we consider prey particles undergoing directed motion with propulsion speed $v_0$,
\begin{equation}
  \dot\x_k = v_0\vec e_k + \sqrt{2D_0}\nois_k, \quad 
  \dot\vhi_k = \sqrt{2/\tx}\eta_k,
  \label{eq:lang}
\end{equation}
where the $\nois_k,\eta_k$ are zero mean and unit-variance Gaussian white noise. Each particle has an orientation $\vec e_k=(\cos\vhi_k,\sin\vhi_k)^T$ described by the angle $\vhi_k$ with the $x$-axis, which undergoes free rotational diffusion with correlation time $\tx$. Again, the effective diffusion coefficient in the frame of reference of the predator is $D$. The orientational correlation time is related through the no-slip boundary condition to the translational diffusion, $1/\tx=3D_0/\sigma_\text{eff}^2$, with effective particle diameter $\sigma_\text{eff}\simeq1.10688\sigma$.

In Fig.~\ref{fig:active}(a) we show numerical density profiles for different speeds. Clearly, increasing the speed leads to a flatter density profile that sharply declines as the absorbing boundary at $r=a$ is approached. The effect of self-propulsion can thus not be captured by an elevated diffusion coefficient alone (as for the mean-square displacement of a free particle~\cite{howse07}). Indeed, Fig.~\ref{fig:active}(b) shows that the density profile for active particles depends on the value of $D$ in contrast to the passive case.

\begin{figure}[t]
	\centering
	\includegraphics{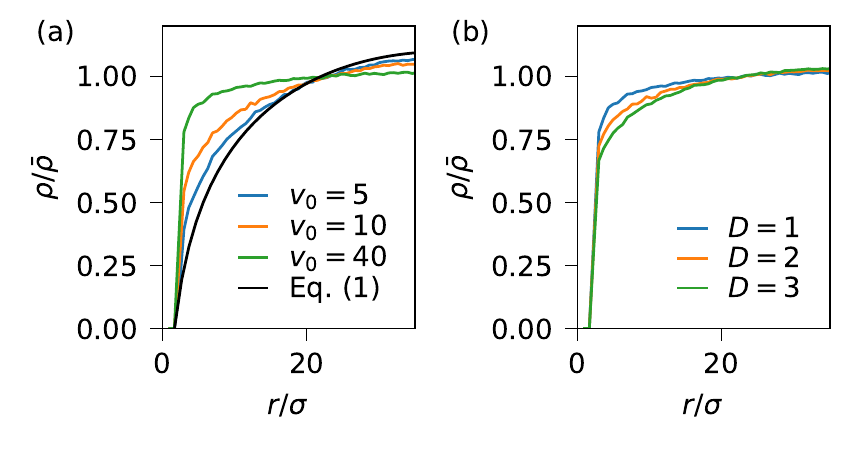}
	\caption{Active diffusion ($L=70$). Density profiles $\rho(r)$ away from the predator for (a)~different speeds $v_0$ at $D=1$ [the black line shows the passive result Eq.~\eqref{eq:rho:p}] and (b)~different $D$ at speed $v_0=10$.}
	\label{fig:active}
\end{figure}

\subsection{Learning optimal behavior} 
\label{sec:learning}

We now assume that each prey particle has limited computational capabilities that allow it to determine a state $s_k\in\mathcal S$ and to perform an action $a_k\in\mathcal A$. Both the possible states $\mathcal S$ and the actions $\mathcal A$ are discrete sets of a few possibilities, which are related through a $Q$-matrix with entries $Q_{sa}$. Instead of integrating Eq.~\eqref{eq:lang}, particles now evolve according to the following scheme with time step $\delta t$:
\begin{enumerate}
	\item determine the state $s_k$ of each particle
	\item determine the action $a_k$ that maximizes $Q_{s_ka_k}$
	\item translate all particles positions
	\begin{equation}
	  \x_k(t+\delta t) = \x_k(t) + v(a_k)\delta t\vec e_k + \sqrt{2D_0\delta t}\nois_k
	  \label{eq:trans}
	\end{equation}
	and orientations
	\begin{equation}
	  \vhi_k(t+\delta t) = \vhi_k(t) + \om(a_k)\delta t + \sqrt{2\delta t/\tx}\eta_k,
	  \label{eq:rot}
	\end{equation}	
\end{enumerate}
which is repeated. The action $a$ thus determines the propulsion speed $v(a)$ and the torque $\om(a)$.

To proceed, we need to determine the $Q$-matrix relating state to action. We break the learning into several episodes, and each episode is divided into multiple steps. Each episode represents a simulation, where at the beginning all prey particles are initialized randomly. The predator remains located in the center of the simulation box throughout the learning process. After 10,000 time steps $\delta t$, we determine the states $s_k$ and each prey particle receives a reward $R_k$. Then the prey particles choose new actions, which are applied for the next learning step. The action policy is based upon the current $Q$-matrix according to an $\eps$-greedy exploration scheme,
\begin{equation}
  a_k = \begin{cases}
    \argmax Q_{s_ka} & \text{with prob. } 1 - \eps_n \\ \text{random action} & \text{with prob. } \eps_n.
  \end{cases}
  \label{eq:greedy}
\end{equation}
At the beginning of the learning process, we start with $\eps_0=1$ and decrease its value according to $\eps_n=0.995^n$ with rising experience of the prey, where $n$ enumerates the learning episodes.

The $Q$-matrix is initialized with all entries set to one. The entry for each state-action pair ($s,a$) is then updated after each step by the rule
\begin{equation}
  Q_{s_ka_k} \leftarrow Q_{s_ka_k} + \alpha_n [R_k + \gamma\max_a Q_{s_k'a} - Q_{s_ka_k}],
\end{equation}
where $s_k'$ is the new state after advancing the simulation with $a_k$. The future reward discount factor is set to $\gamma = 0.99$. The learning rate $\alpha_n=0.8/(0.8+n)$ depends on the episode $n$ and decreases.

\begin{figure}[t]
	\centering
	\includegraphics{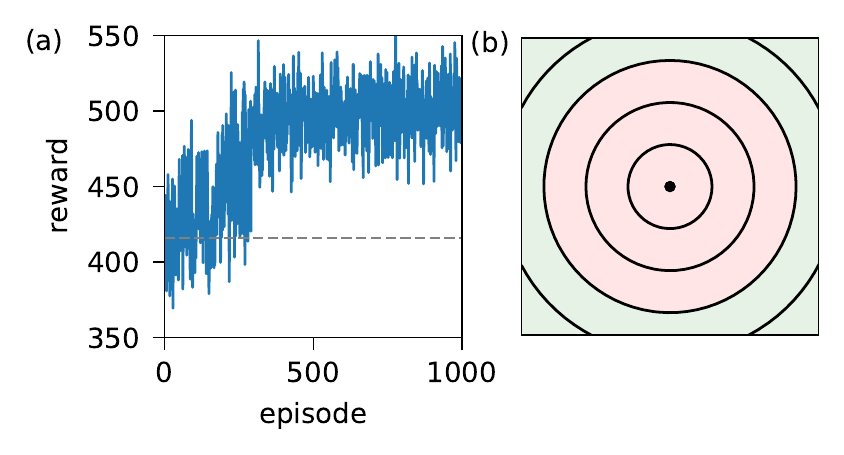}
	\caption{Reinforcement learning. (a)~Performance of the $N=150$ prey particles as learning progresses. Plotted is the average reward per particle $\sum_kR_k/N$ at the end of each episode as a function of episodes. Speed of active particles is $v_0=20$. The gray dashed line indicates the average reward $\mean{R_k}_\text{hom}\approx 416$ for a homogeneous system. (b)~Representation of the $Q$-matrix, whereby each concentric ring corresponds to one discrete state. Particles in the red rings move actively ($v=v_0$), while particles in the outer green rings diffuse passively ($v=0$).}
	\label{fig:reward}
\end{figure}

As a first example, we assume that prey particles can somehow estimate their distance to the predator (\emph{e.g.}, through sensing a chemical signal exuded by the predator~\cite{sengupta11}). The discrete state $s_k=\lfloor(\x_k-\x_\text{p})/\lam\rfloor$ then measures the distance to the position $\x_\text{p}$ of the predator with spacing $\lam$ and floor function $\lfloor\cdot\rfloor$. The reward is calculated as
\begin{equation}
	R_k = |\x_k-\x_\text{p}|^2.
  \label{eq:rew}
\end{equation}
The learning process is performed in a square box with edge $L=50$ and $N=150$ prey particles over $1000$ episodes, each with $20$ learning steps. The actions are restricted to switching the motility on/off,
\begin{equation}
  v(a) = \begin{cases}
    0 & (a=1) \\ v_0 & (a=2)
  \end{cases}
  \label{eq:mot}
\end{equation}
where $v_0=20$. During the learning process, we measure the success through the average reward at the end of each episode. The reward progress is shown in Fig.~\ref{fig:reward}(a). At the beginning of the learning process, the particles are uniformly distributed over the entire box. In this case, the mean-square distance between prey and predator located in the center is
\begin{equation}
	\mean{R_k}_\text{hom} = \frac{1}{A}\IInt{x}{-L/2}{L/2}\IInt{y}{-L/2}{L/2}(x^2+y^2) \approx \frac{L^2}{6},
\end{equation}
which corresponds to a reward of about $416$. It reaches a plateau after about $500$ episodes at an average value of about $500$, which corresponds to an average distance of about $22$ to the predator. The resulting policy for five discrete states is shown in Fig.~\ref{fig:reward}(b), where particles in the red area move actively with speed $v_0$ while particles in the area highlighted in green only undergo diffusive motion. The radius $R_\ast$ of the active region is a non-trivial function of system size $L$, discretization, and speed $v_0$. Due to the periodic boundaries, to maximize the reward the passive region needs to be sufficiently large otherwise prey particles return to the active region too fast.

\begin{figure}[t]
  \centering
  \includegraphics{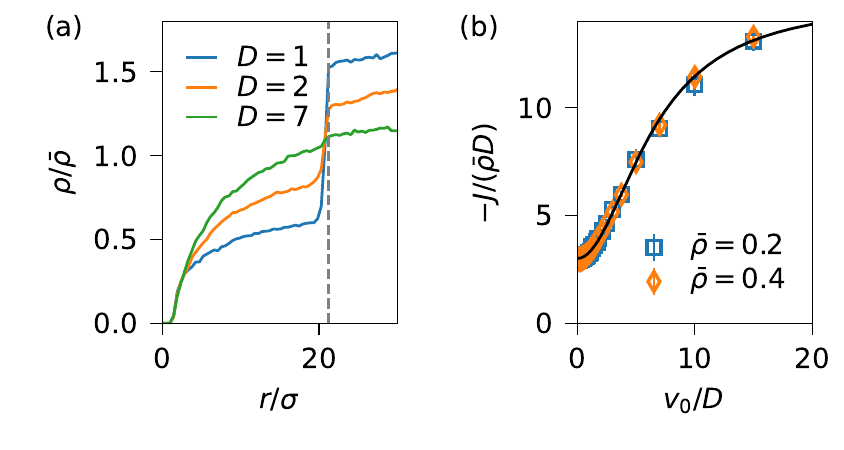}
  \caption{Distance-based strategy. (a)~Numerical density profiles $\rho(r)$ away from the predator for $\bar\rho=0.2$ and different diffusion coefficients $D$ employing the $Q$-matrix with speed $v_0=5$ (different from the learning speed). The dashed line indicates the transition radius $R_\ast\approx21.2$ between passive and active regions, cf. Fig.~\ref{fig:reward}(b). (b)~Reduced current $-J/(\bar\rho D)$ through the absorbing boundary for two global densities and several diffusion coefficients and speeds. The full line shows Eq.~\eqref{eq:J}.}
  \label{fig:optimal}
\end{figure}

After learning the $Q$-matrix at one speed $v_0$ and diffusion coefficient $D$, we perform further simulations with the final $Q$-matrix for different $v_0$ and $D$. These simulations integrate the same equations~\eqref{eq:trans} and~\eqref{eq:rot} of motion but the action is determined deterministically through the rule $a_k=\argmax Q_{s_ka}$. In Fig.~\ref{fig:optimal}(a), we show numerical density profiles $\rho(r)$ at speed $v_0=5$ and global density $\bar\rho_0=0.2$ for different diffusion coefficients $D$. For low diffusion coefficients, we see a density discontinuity at the threshold between active and passive particles since the passive particles accumulate in the outer regions of the box, but there is always a certain amount of motile particles in the active region. The higher the diffusion coefficient, and therefore also the rotational diffusion, the narrower the gap becomes, which disappears above about $D=7$ and the system is dominated by diffusion. In Fig.~\ref{fig:optimal}(b), we show for two densities that the current $-J$ through the absorbing boundary increases with $v_0$ and is always larger than the passive current. On first glance this seems counterintuitive since the prey particles accumulate away from the predator, but the self-propulsion also leads to an increased probability to encounter the predator. Tentatively replacing the passive diffusion coefficient $D$ in Eq.~\eqref{eq:J:p} by an elevated active diffusion leads to the expression
\begin{equation}
  -J = -J_0 + \frac{v_0^2}{2D}\frac{1}{c_1+c_2(v_0/D)^2}\bar\rho,
  \label{eq:J}
\end{equation}
where we assume that for large speeds the current saturates (trajectories through the active region become basically straight lines with a fixed probability to hit the predator). Figure~\ref{fig:optimal}(b) demonstrates that this expression describes the measured current very well with fit parameters $c_1\simeq1.75$ and $c_2\simeq0.042$. Hence, even though the prey successfully accumulate away from the predator, the goal of not getting caught (on average) is not achieved.


\section{Navigating chemical gradients}

\subsection{External gradient}

How can prey particles improve their chances? As already mentioned, one means of communication at the microscale is the release and sensing of signaling molecules, which diffuse quickly and create a gradient. Specifically, let us assume that the predator exudes these signaling molecules with rate $\gamma_\text{c}$ and diffusion coefficient $D_\text{c}$. In principle, predator and prey move much slower than molecules diffuse. In this quasistatic limit, the prey particles effectively move in a concentration field
\begin{equation}
	c(\x) = \frac{\gamma_\text{c}}{4\pi D_\text{c}}  \frac{e^{-|\x-\x_\text{p}|/\lambda}}{|\x-\x_\text{p}|}
	\label{eq:concentration}
\end{equation}
that is parametrized by the position $\x_\text{p}$ of the predator alone. Here we assume that the predator and prey move diffusively in two dimensions close to a substrate, and the concentration profile is that within the semispace above the substrate. A decay of signaling molecules leads to the exponential factor with decay length $\lam$. In the following, we set $\gamma_\text{c}/D_\text{c}=1$ and $\lam=10$. Importantly, through the concentration gradient prey particles can now respond to the orientation of the predator rather than just distance.

We consider two further state-action sets. The first case is that of motility switching with Eq.~\eqref{eq:mot} but now depending on whether the predator is in front or behind them, see Fig.~\ref{fig:policy}(a). Formally, we distinguish these two states through
\begin{equation}
  \mathbf{e}_k\cdot \nabla c|_{\x_k} = \begin{cases}
    \leq 0 & \text{oriented away from} \\ > 0 & \text{oriented toward}
  \end{cases}
\end{equation}
the predator. We perform the learning process over $1000$ episodes, $20$ steps each, and with learning parameters $\alpha$ and $\gamma$ as described in Sec.~\ref{sec:learning} using as reward again the squared distance Eq.~\eqref{eq:rew}. The outcome policy is that particles which are facing away from the predator should move actively in order to increase the distance to the predator. Particles which are oriented towards the predator should remain passive until either the motion of the predator or rotational diffusive motion leads to a change in state.

\begin{figure}[t]
	\centering
	\includegraphics{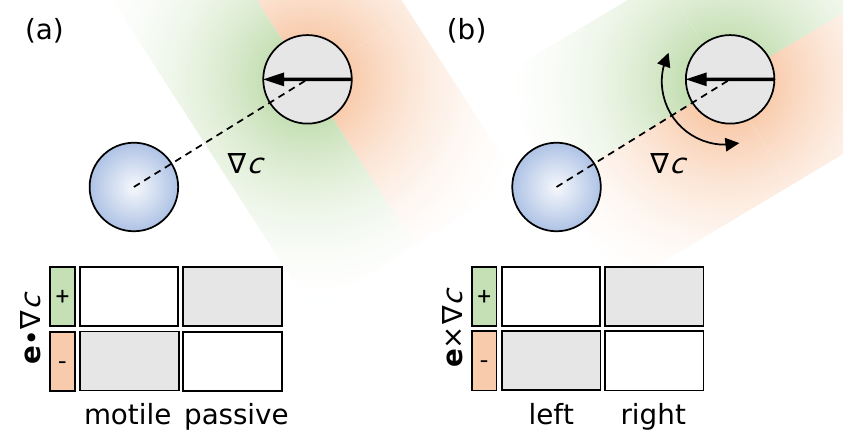}
	\caption{Gradient-based strategies. (a)~Motility switching based on the relative orientation with respect to the concentration gradient $\nabla c$ (pointing towards the predator). The $Q$-matrix is sketched: to maximize the distance, prey particles become motile when facing away and passive when facing towards the predator. (b)~Orientation adaptation. Prey particles are always active and turn away from the predator.}
	\label{fig:policy}
\end{figure}

The second case is an orientation-adaption model, where particles act by either turning themselves to the right or to the left according to
\begin{equation}
  \om(a) = \begin{cases}
    -\om_0 & (a=1) \\ +\om_0 & (a=2)
  \end{cases}
  \label{eq:torque}
\end{equation}
with torque (angular speed) $\om_0=13.3$. The speed $v_0$ in this model is constant. We consider again two possible states, which are sketched in Fig.~\ref{fig:policy}(b). The first one is that the predator is on a particle's left side, the other one is that the predator is on its right side. We can express this with the two-dimensional cross product of the orientation vector and the gradient,
\begin{equation}
  \vec e_k\times\nabla c|_{\x_k} = \begin{cases}
    \leq 0 & \text{predator right} \\ > 0 & \text{predator left}.
  \end{cases}
\end{equation}
After the learning process, the particles follow the policy derived from the $Q$-matrix shown in Fig.~\ref{fig:policy}(b), which indicates that the prey should always turn away from the predator.

As before, we test the quality of the derived policies by measuring the success of a predator in catching prey particles that follows those policies~\cite{sm}. In contrast to the previous simulations, now the predator not only passively catches particles that come below a certain distance threshold, it also follows a fixed chasing strategy: it always focuses on chasing the nearest prey particle with constant speed $v_\text{p}$. The prey particles are still randomly set back into the box to maintain a constant density. We use the resulting current induced by the predator to evaluate both policies. We perform simulations with different speeds $v_0$ of the prey particles while the predator moves with speed $v_\text{p}$ and rotates with torque $\om_\text{p}=\om_0=13.3$ towards the nearest prey particle.

\begin{figure}[t]
	\centering
	\includegraphics{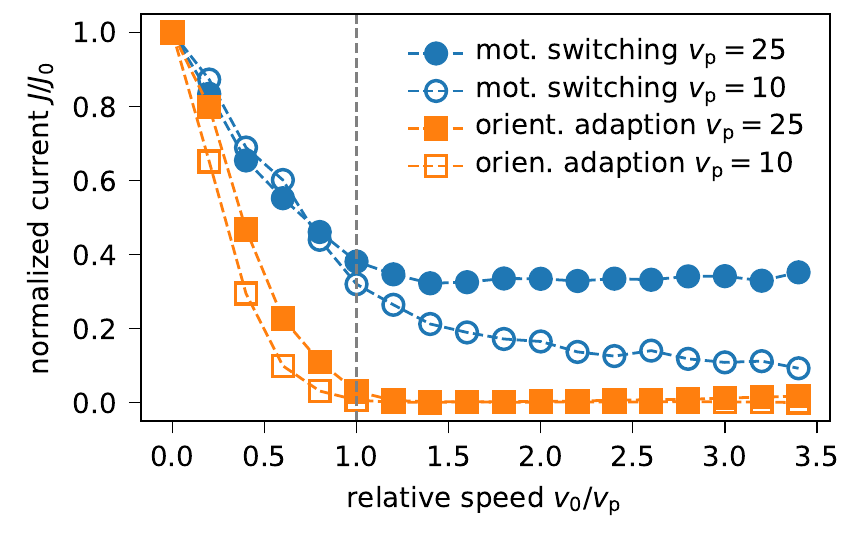}
	\caption{Normalized current of both gradient-based policies as a function of the relative speed $v_0/v_\text{p}$ between predator and prey particles for global density $\bar\rho=0.17$. Shown are results for two predator speeds $v_\text{p}=10$ (open symbols) and $v_\text{p}=25$ (filled symbols). For prey that outruns the predator ($v_0>v_\text{p}$) the current becomes independent of the relative speed and drops to zero for the orientation adaption.}
	\label{fig:comparison_current}
\end{figure}

Figure~\ref{fig:comparison_current} shows the normalized current $J/J_0$ of the two policies depending on the relative speed $v_0/v_\text{p}$ at global density $\bar\rho=0.17$. The current is normalized by the current $J_0$ in a system with only passive prey particles. We observe for both models that the current decreases up to a relative speed of about $v_0\approx v_\text{p}$, \emph{i.e.} prey particles are successful in avoiding the predator. Above that point, the current reaches a non-zero plateau in the motility-switch model for fast predators. In the plateau region, the predator catches only particles that are not actively moving. In the orientation-adaption model the current falls to zero, \emph{i.e.}, all prey particles manage to escape as long as they move faster. The predator is able to catch particles only shortly after initialization, when the prey have to reorient away from the predator.

\subsection{Self-generated gradients}

So far, we have considered independent prey particles that react to an external stimulus, here the predator. We now remove the predator and assume that the particles both exude and sense signaling molecules as in quorum sensing~\cite{brown01}. We aim to find a policy so that particles aggregate into clusters at very low densities. Particles now have an excluded volume that we model through the repulsive Weeks-Chandler-Anderson (WCA) pair potential 
\begin{equation}
  u(r) =\left\{\begin{array}{ll} 4\epsilon\left[\left(\frac{\sigma}{r}\right)^{12}-\left(\frac{\sigma}{r}\right)^6 \right]+\epsilon, & r/\sigma < 2^{1/6}  \\
  0, & r/\sigma \ge 2^{1/6} \end{array}\right.
\end{equation}
with distance $r=|\x_i-\x_j|$ between two particles at positions $\x_i$ and $\x_j$. We employ a potential strength $\epsilon=100k_\text{B}T$, which implies hard-disk-like particles with an effective diameter $\sigma_\text{eff}=1.10688\sigma$~\cite{barker67}.

\begin{figure}[t]
	\centering
	\includegraphics{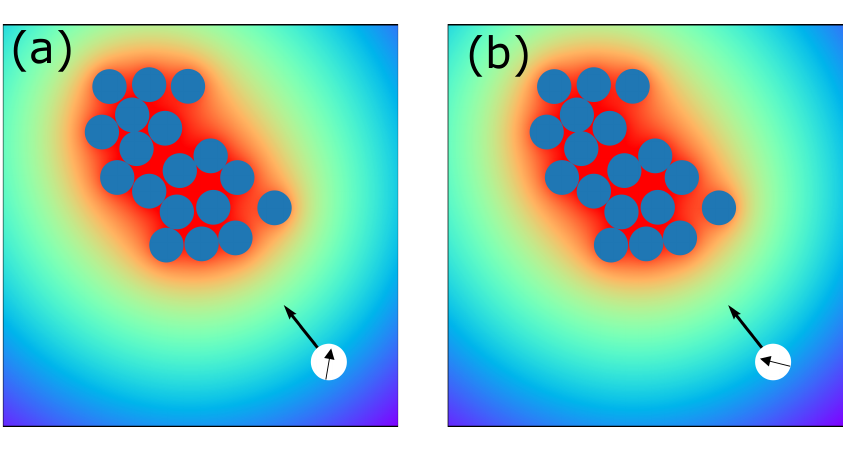}
	\caption{Concentration field $c(\x)$ of signaling molecules produced by the blue particles (red: high $c$; blue: low $c$). The white particle is sensing the concentration gradient with two possible states: the higher density is (a)~to the left or (b)~to the right.}
	\label{fig:multi_states}
\end{figure}

In this model, all particles produce and sense signaling molecules. In regions of high density there is also a high concentration of signaling molecules. We use this concentration as a proxy for how far a particle is away from regions with higher density of particles. We define the reward for each particle as the local concentration 
\begin{equation}
  R_k = c(\x_k) = \frac{\gamma_c}{4\pi D_c} \sum_{i\neq k}^{} \frac{e^{-|\x_k-\x_i|/\lambda}}{|\x_k-\x_i|}
\end{equation}
that it senses. Again, we define two different states similar to the orientation-adaption model, but instead of responding to an external source of signaling molecules, the particles respond to each other. The two possible states are demonstrated in Fig.~\ref{fig:multi_states}. We define the states through the sign of the cross product between the orientation of a particle and the local gradient
\begin{equation}
	\vec e_k\times\nabla c|_{\x_k} = \begin{cases}
	  \leq 0 & \text{higher density right} \\ > 0 & \text{higher density left}.
	\end{cases}
\end{equation}
Depending on the state, the particle can choose to turn left or right, see Eq.~\eqref{eq:torque}.

The learning process is performed with $N=50$ particles in a square system with $L=20$. All parameters of the reinforcement learning algorithm are the same as in the previous examples. We set the active speed to $v_0=20$ and learn over $1000$ episodes with $20$ steps each. The $Q$-matrix results in a policy in which particles align with the gradient and thus orient towards higher concentrations. We then employ the resulting $Q$-matrix to investigate the clustering process depending on the strength of the reaction of the particles to their self-generated chemical field. While similar to Ref.~\citenum{pohl14}, our simulations differ in the following points: First, we consider a torque that is not depending on the actual value of the gradient but only on the sign of of the gradient. Moreover, we do not consider a translational diffusiophoretic motion due to the concentration gradient.

\begin{figure}[t]
	\centering
	\includegraphics{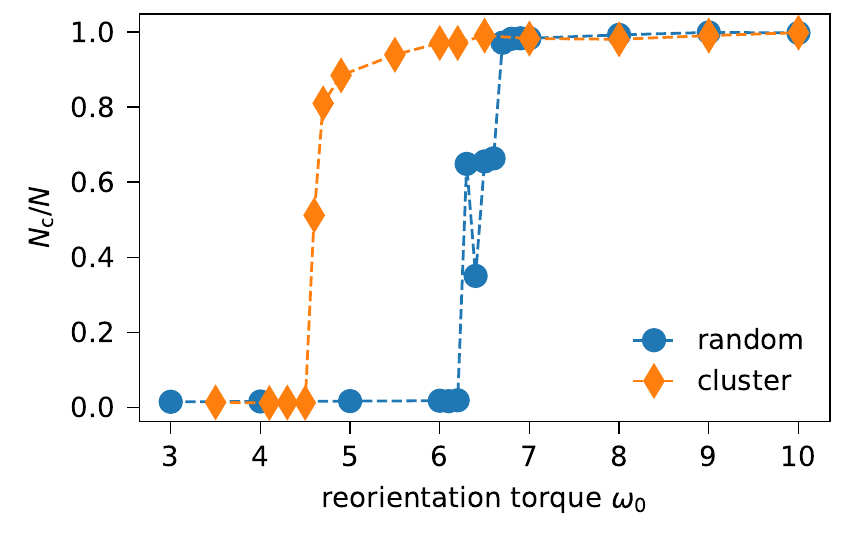}
	\caption{Average fraction $N_\text{c}/N$ of particles in a cluster as a function of reorienting torque $\om_0$ for a system that was initialized either randomly in a homogeneous state or in a single cluster.}
	\label{fig:cluster}
\end{figure}

We investigate dilute systems with packing fraction $\phi=N\pi(\sigma_\text{eff}/2)^2/(2L)^2\simeq0.05$ and simulate a total of $N=255$ particles. We choose an active speed $v_0=60$ corresponding to a Péclet number $\text{Pe}=\hat v_0/\sqrt{2D_0/\tx}\simeq27$ (with dimensionful speed $\hat v_0$). Figure~\ref{fig:cluster} shows that for small reorientation torque $\omega_0$ the system is an active homogeneous gas while for large $\omega_0$ all particles collapse into one single large cluster. We consider a cluster as an assembly of $N_\text{c}\geq2$ particles. A cluster is determined by all particles that are mutually ``bonded'' (\emph{i.e.}, they are within the cut-off radius of the interaction potential). In the intermediate regime, we study the system from two initial configurations: either we start the simulation with all particle positions initialized randomly or we start with a single cluster. Figure~\ref{fig:cluster} shows that there is considerable hysteresis for our small system and that the formation of the cluster from the homogeneous state occurs though nucleation with small clusters decaying. In contrast, once a large cluster has formed it is stable down to small values of $\om_0$.


\section{Conclusions}

To summarize, we have extended the model of active Brownian particles to speeds and torques that depend on some discrete action $a_k\in A$, see Eqs.~\eqref{eq:trans} and~\eqref{eq:rot}. This action is chosen through a $Q$-matrix, the maximal entry of which for a given state $s_k\in S$ determines $a_k$. This $Q$-matrix can be determined through well-established algorithms known as reinforcement learning, which require as further input a reward function that evaluates the ``utility'' of the current state to the system. We have verified this approach for two toy models. First, we have studied prey particles reacting to a predator, where the reward is given by the distance of the prey from the predator. Particles don't interact directly but only through the predator, aggregating into (in a periodic system) domains away from the predator. Still, we found strong differences between state-action pairs with respect to the success avoiding the predator (measured as a current through an absorbing boundary). These different state-action pairs represent the information that might be available and possible actions. Here we have focused on the local concentration of some signaling molecules, but other cues like light~\cite{jekely08}, gravity~\cite{hagen14,colabrese17}, viscosity~\cite{liebchen18,datt19} \emph{etc.} leading to the different $x$-taxis might be used. Alternatively, run-and-tumble bacteria like \emph{E. coli} might adopt their tumbling rate. It will also be interesting to consider the influence of unavoidable concentration fluctuations of the signaling molecules on optimal search strategies~\cite{kromer20}. Second, we have considered as reward the local concentration gradient. To achieve aggregation, the reorientation torque needs to overcome the rotational diffusion. Our framework can easily be extended to learn the interactions underlying more complex collective behavior.


\begin{acknowledgments}
	We acknowledge financial support through the Emergent AI Center funded by the Carl-Zeiss-Stiftung. We thank Michael Wand for useful discussions.
\end{acknowledgments}

%

\end{document}